\documentclass[useAMS,usenatbib,usegraphicx]{mn2e}
\usepackage{amsmath,aas_macros} % with this, one can use ADS bibtex entries in MNRAS

\newcommand{\lra}[1]{{ \left( #1 \right) }}

\newcommand{\fil}[1]{\langle #1 \rangle}

\newcommand{\pd}[1]{\frac{\partial}{\partial #1}}

\newcommand{\DD}{\ensuremath{\mathrm{D}}}

\voffset =- .6in %for Arxiv use

\title[Turbulence production and pressure support in the IGM]{Turbulence
  production and turbulent pressure support in the intergalactic medium} 

\author[L. Iapichino, W. Schmidt, J. C. Niemeyer and
J. Merklein]{L. Iapichino$^{1}$\thanks{E-mail:
    luigi@ita.uni-heidelberg.de 
}, W. Schmidt$^{2}$, J. C. Niemeyer$^{2}$ and J. Merklein$^{3}$\\
$^{1}$Zentrum f\"ur Astronomie der Universit\"at Heidelberg, 
Institut f\"ur Theoretische Astrophysik, Albert-Ueberle-Str.~2, D-69120 \\
Heidelberg, Germany\\
$^{2}$Institut f\"ur Astrophysik, Universit\"at
  G\"ottingen, Friedrich-Hund-Platz 1, D-37077 G\"ottingen, Germany\\
$^{3}$Abteilung Bioklimatologie, Universit\"at
  G\"ottingen, B\"usgenweg 2, D-37077 G\"ottingen, Germany}

\begin{document}

\date{Accepted 2011 February 14. Received 2011 February 14; in original form 2010 November 30}

\pagerange{\pageref{firstpage}--\pageref{lastpage}} \pubyear{2011}

\maketitle

\label{firstpage}

\begin{abstract}
The injection and evolution of turbulence in the intergalactic medium is studied by means of mesh-based hydrodynamical simulations, including a subgrid scale (SGS) model for small-scale unresolved turbulence. The simulations show that the production of turbulence has a different redshift dependence in the intracluster medium (ICM) and the warm-hot intergalactic medium (WHIM). We show that turbulence in the ICM is produced chiefly by merger-induced shear flows, whereas the production in the WHIM is dominated by shock interactions. Secondly, the effect of dynamical pressure support on the gravitational contraction has been studied. This turbulent support is stronger in the WHIM gas at baryon overdensities $1 \la \delta \la 100$, and less relevant for the ICM. Although the relative mass fraction of the gas with large vorticity is considerable ($52\%$ in the ICM), we find that for only about $10\%$ in mass this is dynamically relevant, namely not associated to an equally large thermal pressure support. According to this result, a significant non-thermal pressure support counteracting the gravitational contraction is a localised characteristic in the cosmic flow, rather than a widespread feature.  
 
\end{abstract}


\section{Introduction}
\label{intro}

Often considered just a
by-product of the virialisation mechanism, the injection of turbulence
energy in the intergalactic medium (IGM) has been recognised as an
interesting process in its own right. Put in a broader context, the
information about non-thermal phenomena inside and outside galaxy
clusters can complement the 
information derived from the hot,
X-ray emitting gas. Turbulence, in particular, is of utmost importance
because of the apparent link to the acceleration mechanisms of cosmic
rays (CR) and to cluster diffuse radio emission
\citep{fgs08,c09,b09,bcd09,ceg10}. Turbulent velocity fluctuations are
also considered to be important for amplifying magnetic fields in the
intracluster medium (ICM) \citep{ssh06} and in the IGM \citep{rkc08}. 

To date, constraints on the turbulent velocity in clusters have been
determined by measuring resonant scattering suppression
\citep{cfj04,wzc09} and by {\it XMM-Newton} spectroscopic observations
of clusters with a compact core \citep{sfs10,sfs10b}. 
Recently, the {\it Suzaku} satellite made deep X-ray observations out
to the virial radius of several clusters possible
\citep{rhz09,gfs09,bms09,hps10,kou10}, thanks to its low and stable
particle background. In most of the 
published data, the observed
clusters show departures from the hydrostatic equilibrium in their
outskirts, in one case \citep{kou10} the non-thermal support is as
large as about $50\%$ of the total pressure\footnote{By `non-thermal pressure' one indicates generically pressure contributions from
  turbulence, magnetic fields and non-thermal particles (cosmic rays);
  in the following, the latter two contributions will not be
  addressed. Moreover, non-equipartition effects of the low-density
  plasma in the cluster outskirts could mimic a non-thermal
  contribution (see \citealt{ws09,rn09}, and references therein).}.  

As for the gas outside clusters, the strongest recent arguments in favour 
of a significant effect of small-scale turbulence have been
derived from low-$z$ O VI observations compared with numerical simulations
\citep{2009MNRAS.395.1875O}. In this work, turbulence had to
be added by hand as a post-process correction since a
subgrid-scale turbulence model was unavailable.  

Whereas the theoretical and numerical exploration of  turbulence in the hot
ICM of galaxy clusters has received a fair
amount of attention (e.g., \citealt{dvb05, vtc06,
  ssh06, sb08, in08,2009MNRAS.398..548B, vbk09, mis09}), the
turbulent state of the cluster outskirts and of the warm-hot
intergalactic medium (WHIM), which is believed to contain 
a significant fraction of the baryons in the low-$z$ universe
\citep{co99, 2008MNRAS.387..577O}, 
has not yet been addressed in a similarly systematic way. The
properties of turbulence in the cluster outskirts have been studied
recently in simulations by \citet{bso10}, who found that those regions
are not in hydrostatic equilibrium, and have a substantial turbulent
pressure support. Based on a sample of sixteen cluster simulations,
\citet{lkn09} conclude that the support of turbulent motions increases
towards the cluster periphery.  \citet{snb10} find that the previous
results lead to a significant reduction of the Sunyaev-Zel'dovich
power spectrum at angular scales of a few arcminutes. \citet{zff10}
present an instructive analysis of the resolved vorticity field and
the influence of the turbulent pressure in cosmological
simulation. Using the dynamical equation for the rate of change of the
divergence \citep[cf.~][]{schmidt09a}, they estimate the effect of
turbulent pressure on the gravitational contraction  
of the baryonic gas. In this way, the role of gas turbulence in the
clustering process is studied (see also \citealt{bpp92}). 

There are several mechanisms that are potentially able to stir the
baryons and inject turbulence in the fluid. In the framework of our
simulations we neglect the 
effects of galaxy motions in the ICM
\citep{bd89,k07,pqs10,ro10} and outflows from AGN activity
\citep{hby06,ss06a,bsh09}. Cluster mergers and curved shocks thus
remain as the main stirring agents to be considered.  

According to the hierarchical scenario for clustering, a halo accretes
most of its mass by mergers. In particular, one can distinguish
between an earlier phase of major mergers, where the ratio of the mass
of the merging subclumps is close to unity, and a subsequent minor
merger phase, when smaller subhaloes fall into the cluster gravitational
well. 

Both mergers phases perturb the cluster medium and are thus related to
the injection of turbulence in the ICM. 
According to \citet{ssh06}, turbulence produced in the major merger
phase has a large volume filling factor, as expected from events which
deeply stir and rearrange the cluster structure
\citep{rbl93,rlb97,rs01,mmb09,pim10}. In case of minor mergers, as the study
of idealised setups has shown, the shear at the boundary between the
ICM and the accreting subcluster triggers the Kelvin-Helmholtz
instability, locally injecting turbulence in the wake of the moving
subclumps \citep{hcf03,ias08,mis09}.  

Another stirring mechanism is linked to the baroclinic vorticity
generations at curved shocks. As known from theory, vorticity $\omega$
is produced where the pressure and density gradients are not
parallel. Taking the curl of the Euler equation \citep{ll59,krc07}, 
\begin{equation}
\label{vort}
\frac{\partial \bomega}{\partial t} = \bmath{\nabla} \times (\bmath{v}
\times \bomega) - \frac{\bmath{\nabla} p \times \bmath{\nabla}
  \rho}{\rho^2} 
\end{equation}
where the second term at the right-hand side is non-vanishing at the locations of non-planar shocks.
In filaments and cluster outskirts the unprocessed gas is accreted
from the voids and accelerated towards the growing structures, where
accretion shocks are formed. The injection of turbulence is therefore
a by-product of the gas accretion at curved shocks.  

The main difference in the two distinct mechanisms described above is
in the driving: in merger events, turbulence is generated by shearing
instabilities, whereas at shocks the generation is driven by
compressional modes. This difference is expected to affect the flow
features  in a quantifiable way that will be explored through our
numerical simulations. Even for a fixed temperature, numerical studies
of forced supersonic turbulence indicate significant differences in
the distributions of density and velocity fluctuations depending on
the forcing \citep{SchmFeder08,Federrath2008,sfh09,frk09}. 

It is clear from the above that turbulence is generated on essentially
all cosmologically relevant scales. Moreover, the turbulent WHIM and
ICM include a temperature range from $10^5 - 10^8$ K and
corresponding Mach numbers ranging from $10^{-2}$ to
transonic values. Adding to the inherent difficulties of
simulating turbulence even under more simplified (i.e., homogeneous
and isothermal) conditions, these complications make the exploration
of large-scale structure turbulence challenging, to say the least. A
convincing statistical analysis of the turbulence properties of the
IGM on numerically resolved scales is infeasible with present codes
and resources.  

 On the other hand, much can be learnt already from looking at the
 magnitude of the production terms and keeping track of the amount of
 turbulent kinetic energy on unresolved scales by means of a
 subgrid-scale (SGS) model. The central simplifying assumption here is
 that turbulence can be considered to be statistically
 isotropic on sufficiently small scales. Clearly, this is only a first
 approximation which cannot replace the information gained by
 increased resolution. However, extensive experience with large-eddy
 simulations (LES), as this technique is commonly referred to, shows
 that many properties of unresolved turbulence (most importantly,
 those related to transport and dissipation) can be captured with a
 certain degree of confidence (for further references,
 see \citealt{snh06,schmidt09}). Once a reliable model for small-scale
 turbulence in cosmological simulations is available, many of the
 questions listed above can already be addressed to some extent. This
 is the approach we will take in this work. 

In a previous paper \citep{mis09}, we introduced a version of LES
suitable for adaptive mesh refinement (AMR) called {\sc fearless}. Its
main properties will be summarised in Section \ref{tools}. Here, we apply
this model to a large-scale structure simulation with adiabatic gas
dynamics for the first time. Considering only the SGS turbulence
energy as a probe for the production of turbulence for the reasons
explained above, we find that its evolution indeed differs
significantly for the WHIM and ICM phases of the IGM. Although this
result is limited by the approximate nature of the turbulence SGS model, we
offer an interpretation in terms of different dominant
production mechanisms in these phases and present supporting evidence
in Section \ref{simulations}.

In Section \ref{turb_press}, we will further elaborate on this approach.
The turbulence SGS model allows us to predict the contribution of the turbulent pressure on the 
grid scale in addition to the effects caused by numerically resolved turbulence. 
Comparing the WHIM and the ICM, we find that the support of the gas against gravitational
contraction by turbulence is more pronounced at low densities in the WHIM than at
higher densities in the WHIM and ICM, in which the support is mainly
thermal. A further important finding which will be discussed is that
the turbulence-supported gas has a fairly low mass and volume
fraction.  

We conclude with a summary of our results and suggestions for future
directions in Section \ref{conclusions}.

\section{Numerical tools}
\label{tools}

This work is based on hydrodynamical simulations of the evolution of
the cosmic large-scale structure, performed using the {\sc 
fearless} numerical technique (Fluid mEchanics with Adaptively Refined
Large Eddy SimulationS; \citealt{mis09}) for simulating intermittent
turbulent flows in clumped media. This tool has been implemented on the
public release of the grid-based, adaptive mesh refinement (AMR) hybrid
(N-Body plus hydrodynamical) code
{\sc enzo} (v.~1.0) \citep{obb05}\footnote{{\sc enzo} homepage:
http://lca.ucsd.edu/portal/software/enzo}.

\subsection{Setup of the simulations}
\label{setup}

A flat $\Lambda$CDM cosmology is assumed, with $\Omega_\rmn{\Lambda} =
0.721$, $\Omega_\rmn{m} = 0.279$, $\Omega_\rmn{b} = 0.046$, $h = 0.7$,
$\sigma_8 = 0.817$, and $n=0.96$.  

The computational box has a side of $100\ \rmn{Mpc}\ h^{-1}$ and is
resolved with a root grid of $128^3$ cells and $128^3$ N-body
particles. The mesh is refined with four additional AMR levels
(refinement factor $N = 2$), leading to the effective spatial
resolution of $l_{\Delta, 4} = 48.8\ \rmn{kpc}\ h^{-1}$. The force resolution of the gravity solver is of the order of $2 \times l_{\Delta, 4}$. The AMR criteria are based on baryon and DM overdensity, with overdensity factors $f = 4$
\citep{in08}. 

The initial redshift of the simulations is $z = 60$, and the initial
conditions are produced with the \citet{eh99} transfer function. The
evolution is then followed to $z = 0$. Additional physics such as
cooling, feedback and 
transport processes is neglected. An ideal equation of state was used
for the gas, with $\gamma = 5/3$.

\subsection{Subgrid-scale model and the {\sc fearless} approach}
\label{sgs}

{\sc fearless} combines AMR with a subgrid scale (SGS) model for the
unresolved turbulence energy, which encompasses the production, the
diffusion and the dissipation of kinetic energy on subgrid scales
\citep[see][]{snh06}. Details, numerical tests and applications to the physics of galaxy clusters are presented elsewhere \citep{mis09}, and here we recall the main features of this tool.

In \citet{snh06} it is shown how the governing equations of a compressible, viscous, self-gravitating fluid can be decomposed into a large-scale (resolved) and a small-scale (unresolved) part by exploiting the \citet{Germano1992} filtering formalism, applied to density-weighted variables \citep{Favre1969}. According to this formalism, once a filtering length is set, a variable $f$ can be decomposed in a smoothed part $\fil{f}$ and a fluctuating part $f'$, with $\fil{f}$ varying only on scales larger than the filter length scale. A filtered quantity $\hat{f}$ is thus defined by
\begin{equation}
\fil{\rho f} = \fil{\rho} \hat{f} \Rightarrow \hat{f}=\frac{\fil{\rho f}}{\fil{\rho}}\,\,.
\end{equation}
In the following, we assume that the filter length scale is generically given by the grid scale $l_{\Delta}$, i.~e., the size of the grid cells at any level
of refinement.

By applying this formalism one can derive the filtered equations of the fluid dynamics (cf.~\citealt{snh06}). For the sake of conceptual clarity, we do not include the cosmological expansion here
and refer the reader to \citet{mis09} for a complete formulation in co-moving coordinates. The
resulting equations read
\begin{align}
\pd{t}\fil{\rho} &+ \pd{r_j}\hat{v}_j\fil{\rho} =\ 0, \label{eq:filmsum}\\
\scriptstyle
\begin{split}
\pd{t}\fil{\rho}\hat{v}_i &+ \pd{r_j}\hat{v}_j\fil{\rho}\hat{v}_i =
-\pd{r_i}\fil{p}+\pd{r_j}\fil{\sigma'_{ij}}\\
&+\fil{\rho}\hat{g}_i-\pd{r_j}\hat{\tau}(v_i,v_j)\,\,,
\end{split}\label{eq:filmomsum}
\\
\begin{split}
\pd{t}\fil{\rho}e_{\mathrm{res}} &+\pd{r_j}\hat{v}_j\fil{\rho}e_{\mathrm{res}}=
-\pd{r_i}\hat{v}_i\fil{p}+\pd{r_j}\hat{v}_i\fil{\sigma'_{ij}}\\
&+\fil{\rho}(\lambda+\epsilon)-\hat{v}_i\pd{r_j}\hat{\tau}(v_i,v_j)\\
&+\fil{\rho}\hat{v}_i\hat{g}_i-\pd{r_j}\hat{\tau}(v_j,e_{\mathrm{int}})\,\,,
\end{split}\label{eq:filresetotsum}
\end{align}
where $\rho(r_i,t)$ is the baryon density, $v_i(r_i,t)$ are the velocity components and $e(r_i,t)$ is the total specific energy,  $p$ the pressure, $g_i$ the gravitational acceleration and
$\sigma'_{ij}$ the viscous stress tensor. In the above equations, the generalised moments of arbitrary quantities $f$ and $g$ are given by
\begin{align}
\hat{\tau}(f,g)=&\ \fil{\rho f g} - \fil{\rho} \hat{f} \hat {g}.
\end{align}

The filtering of energy leads to the definition of a total resolved energy $e_{\mathrm{res}}=e_{\mathrm{int}}+ 1/2\ \hat{v_i}\hat{v_i}$,where $e_{\mathrm{int}}$ is the internal energy, and the term $1/2\ \hat{v_i}\hat{v_i}$ is the resolved kinetic energy. On the other hand, the filtered kinetic energy $\hat{e}_{\mathrm{kin}}$ includes also an unresolved contribution, expressed by a second-order moment of the velocity field $\hat{\tau}(v_i,v_j)$, the turbulent stress tensor:
\begin{equation}
\hat{e}_{\mathrm{kin}} = \frac{1}{2}\hat{v_i}\hat{v_i} +  \frac{1}{2}\hat{\tau}(v_i,v_j)/ \fil{\rho}
\label{ekin}
\end{equation}
As in \citet{Germano1992} we identify the trace of $\hat{\tau}(v_i,v_j) / \fil{\rho}$ with the square of the SGS turbulence velocity $q$, so that we define the SGS turbulence energy as
\begin{equation}
e_{\mathrm{t}}=\frac{1}{2}q^{2}:= \frac{1}{2}\hat{\tau}(v_i,v_i)/\fil{\rho}.
\end{equation}
Since the trace of $\hat{\tau}(v_i,v_j)$ can be added to the thermal pressure $\fil{p}$ in the filtered momentum equation~(\ref{eq:filmomsum}),
this identity immediately implies that
\begin{equation}
	p_{\mathrm{t}}=\frac{2}{3}\fil{\rho}e_{\mathrm{t}}
\end{equation}
is the turbulent pressure associated with the turbulent
velocity fluctuations on length scales smaller than the grid scale $l_{\Delta}$.

The governing equation of $e_{\mathrm{t}}$, as derived by the filtering of the equations of fluid dynamics, is
\begin{equation}
\pd{t}\fil{\rho}e_{\mathrm{t}}+\pd{r_j}\hat{v}_j
\fil{\rho}e_{\mathrm{t}}=\ \mathcal{D}+\Sigma+\Gamma-\fil{\rho}(\lambda+\epsilon)\,\,,
\label{eq:etsum}
\end{equation}
where the terms at the right-hand side have to be explicitly defined as a function of large-scale filtered quantities and of $e_{\rmn{t}}$. Their definitions (the so-called closures) represent the turbulence SGS model. % and are presented in detail in \citet{mis09}. 
The physical interpretation and the expressions of the terms in equation (\ref{eq:etsum}) are the following:
\begin{itemize}
\item $\mathcal{D}$ represents the diffusion of SGS turbulence energy. Its expression is based on the gradient-diffusion hypothesis \citep{Sagaut2006}
\begin{equation}
\label{eq:trans}
\mathcal{D}=\pd{r_i}C_{\mathcal{D}}\fil{\rho} l_{\Delta} q^2 \pd{r_i} q\,\,,
\end{equation}
with $C_{\mathcal{D}} = 0.4$, as inferred by numerical experiments \citep{snh06}, and $l_{\Delta}$ is the cutoff scale.
\item $\Sigma$ is the production term, i.e.~the term which accounts for the flux of kinetic energy from the resolved to the SGS component. The production terms arises in equation (\ref{eq:etsum}) as the contraction of the turbulent stress tensor and the Jacobian of the resolved velocity:
\begin{equation}
\Sigma = -\hat{\tau}(v_i, v_j)\frac{\partial \hat{v}_i}{\partial r_j}.
\end{equation}
The turbulence stresses $\hat{\tau}(v_i, v_j)$ are given by the commonly used eddy-viscosity closure. In the compressible formulation \citep{snh06,mis09}, this closure reads
\begin{equation}
	\hat{\tau}(v_i,v_j)=-2\fil{\rho}C_{\nu} l_{\Delta} q
	S^*_{ij}+\frac{1}{3}\delta_{ij}\fil{\rho}q^2.
\end{equation}
where $S_{ij}^{\ast}$ is the trace-free part of the rate-of-strain tensor, 
\begin{equation}
	S_{ij}=\frac{1}{2}\left(\frac{\partial \hat{v}_i}{\partial r_j} +  \frac{\partial \hat{v}_j}{\partial r_i}\right).
\end{equation}
\item $\Gamma$ accounts for small-scale gravitational effects, and is neglected in our implementation. As further simplifications in the model, the influence of the viscous stress tensor  $\fil{\sigma'_{ij}}$ in equations (\ref{eq:filmomsum}) and (\ref{eq:filresetotsum}) is neglected (which is well justified for high Reynolds numbers), and so is also the SGS transport of internal energy, given by the divergence of $\hat{\tau}(v_j,e_{\mathrm{int}})$ in equation (\ref{eq:filresetotsum});
\item $\epsilon$ is the term for SGS turbulence dissipation into internal energy, treated as an effect acting only at subgrid scale. This term is therefore added to the numerical dissipation of the code, but is modelled in a physically motivated way (see \citealt{mis09} for its role in the physics of the ICM). Its closure follows \citet{Sarkar1992}:
\begin{equation}
\label{sarkar1}
\epsilon=C_{\epsilon}\frac{q^3}{l_{\Delta}}(1+\alpha_1 M_{\mathrm{t}}^2)
\end{equation}
where $C_{\epsilon} = 0.5$ \citep{Sagaut2006}, $\alpha_1 = 0.5$ and $M_{\rmn{t}} = q / c_{\rmn{s}}$ is the turbulent Mach number, where $c_{\rmn{s}}$ is the speed of sound.
\item $\lambda$ is the pressure dilatation term, which models the effect of unresolved pressure fluctuations and, effectively, is an exchange term between $e_{\rmn{t}}$ and $e_{\rmn{int}}$. Like for the previous term, a closure suggested by \citet{Sarkar1992} is adopted \citep{mis09}.  
\end{itemize}
An enlightening graphical representation of the SGS terms and their role in the energy budget is given by \citet[][cf.~\citealt{maier08}]{ims09}.

The innovation of the {\sc fearless}
approach is a consistent treatment of the interchange between the
kinetic energy on resolved length scales and the SGS turbulence energy
for a cutoff length that varies
in space and time. Apart from the energy flux through the turbulent
cascade, this involves the increase of resolved kinetic energy at the
cost of SGS turbulence energy if refined subgrids are inserted and,
thus, the numerical cutoff length scale decreases. 

The interplay between the AMR of {\sc enzo} (based on the method of \citealt{bc89}) and the turbulence SGS model in {\sc fearless} exploits an additional assumption of Kolmogorov scaling of the turbulent energy \citep{k41,f95}. Considering two AMR levels with spatial resolutions $l_{\Delta,i}$
and $l_{\Delta,j}$, it means that, at grid refinement (or derefinement), the SGS turbulence energies are statistically related by 
\begin{equation}
\frac{e_{\mathrm{t},i}}{e_{\mathrm{t},j}}=\frac{q_i^2}{q_j^2}\sim
\lra{\frac{l_{\Delta,i}}{l_{\Delta,j}}}^{2/3}\,\,.
\label{eq:qscale}
\end{equation}
When a region is refined, a new ``grid'' is created, i.e. the refined grid patch is handled as a single AMR object. In this new fine grid the values of the hydrodynamical variables are interpolated from the coarse grid,
 but the SGS turbulence energy is scaled according to equation (\ref{eq:qscale}), and the velocities are corrected such that the sum of resolved energy and turbulent energy remains conserved. An opposite procedure applies to grid derefinement, as described by \citet{mis09}.

An important preliminary check concerns the range of applicability of the turbulence SGS model, which is devised to the study of, at most, moderately compressible flows. This limitation is incorporated in the code as a safeguarding mechanism for the value of the turbulent Mach number. In order to prevent numerical instabilities, a threshold is set at $M_{\rmn{t, max}} = \sqrt{2}$, as motivated in \citet{mis09}. 

\begin{figure}
  \resizebox{\hsize}{!}{\includegraphics{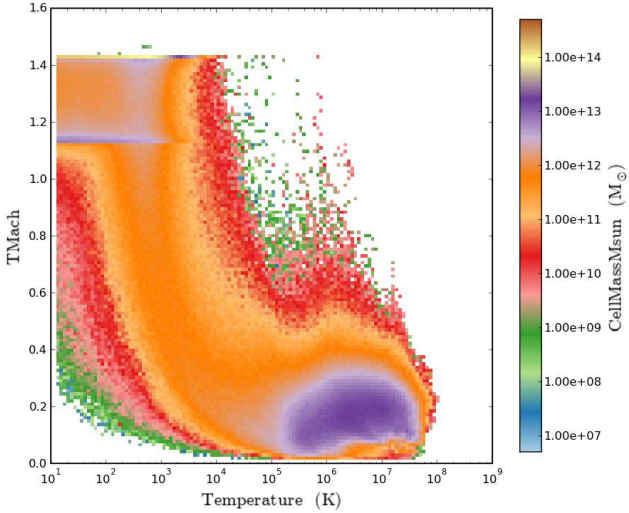}}
  \caption{Two-dimensional mass distribution function of the turbulent Mach number $M_{\rmn{t}}$ as a function of the gas temperature, at the redshift $z = 0$. The mass is coded according to the colour bar on the right. The sharp horizontal line at $M_{\rmn{t}} = 1.41$ is caused by the safeguarding mechanism in {\sc fearless}. A second horizontal line, visible at $M_{\rmn{t}} = 1.41 \times 2^{-1/3} = 1.12$ and $T < 10^4\ \rmn{K}$, is an artifact produced by the interaction of the threshold on $M_{\rmn{t}}$ and the refinement at the first AMR level.} 
  \label{tmach-t}
\end{figure}

The issue of the applicability of the turbulence SGS model for different baryon temperatures is summarised in Fig.~\ref{tmach-t}. Most of the cold gas ($T < 10^4\ \rmn{K}$) lies in the plot at $M_{\rmn{t}} = \sqrt{2}$: for this baryon phase the low sound speed makes the gas motions very supersonic. The turbulence production is therefore unphysical for this gas, because the implemented SGS model is not suitable for its study, and also because the cold and rarefied medium is poorly resolved by our refinement criteria. Anyway, the study of this cold gas phase is physically not well posed in simulations without UV background heating. 

In order to circumvent this shortcoming, we will limit our analysis to the gas with $T > 10^5\ \rmn{K}$. The threshold on the turbulent Mach number does not affect the gas above this temperature, where $M_{\rmn{t}}$ is typically below unity. The SGS turbulent energy is thus not a dominant component in the energy budget, as already discussed in \citet{mis09}.

\section{Evolution of the WHIM and the ICM}
\label{simulations}

\subsection{Baryon phases}
\label{phases}

\begin{figure}
  \resizebox{\hsize}{!}{\includegraphics{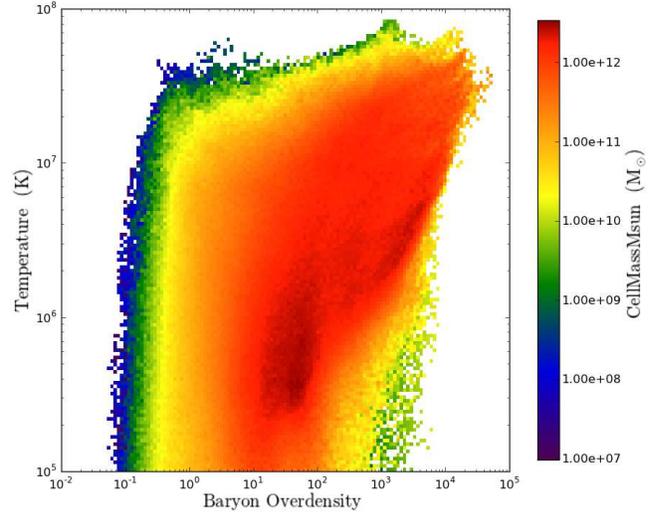}}
  \caption{Two-dimensional mass distribution function of the gas temperature as a function of the baryon overdensity, at $z = 0$. The mass is coded according to the colour bar on the right.} 
  \label{t-rho}
\end{figure}

The main results of this work make use of a distinction of the baryons in two phases. This distinction is somewhat arbitrary, and indeed the mass distribution function in Fig.~\ref{t-rho} shows that the cosmic gas is characterised by a continuum of gas states in temperature and density. The problem is generally addressed in the literature by using a threshold based on gas temperature \citep[e.g.,][]{co99} and/or density \citep{vbg09,soh08}. Ideally, a physically motivated distinction should be done on a dynamical basis, between gas belonging to virialised structures or not. This criterion would be computationally too demanding, and is not used in our analysis.

In the following we will make a distinction based on the baryon overdensity of the gas $\delta = \rho / (\Omega_{\rmn b} \rho_{\rmn{cr}})$, where $\rho_{\rmn{cr}} = 3 H_0^2 /( 8 \pi G)\ (1 + z)^3$ is the critical density at redshift $z$. The gas will be labelled as `WHIM' if $\delta < 10^3$, and as `ICM' if  $\delta > 10^3$. As discussed at the end of Section \ref{sgs}, for both phases the additional constraint $T > 10^5\ \rmn{K}$ is imposed. The former baryon phase is mostly to be found in filaments and in outer halo atmospheres, whereas the latter is in the potential wells of groups and clusters (cf.~Fig.~\ref{panels}). From the definition of the AMR criteria in our simulation, it follows that the ICM gas is always resolved at the highest AMR level $l_{\Delta} = 4$, whereas the mass-weighted average refinement level for the WHIM is 2.6.

\subsection{Turbulent energy}
\label{results}

\begin{figure*}
\centering
\includegraphics[width=0.9\linewidth,clip]{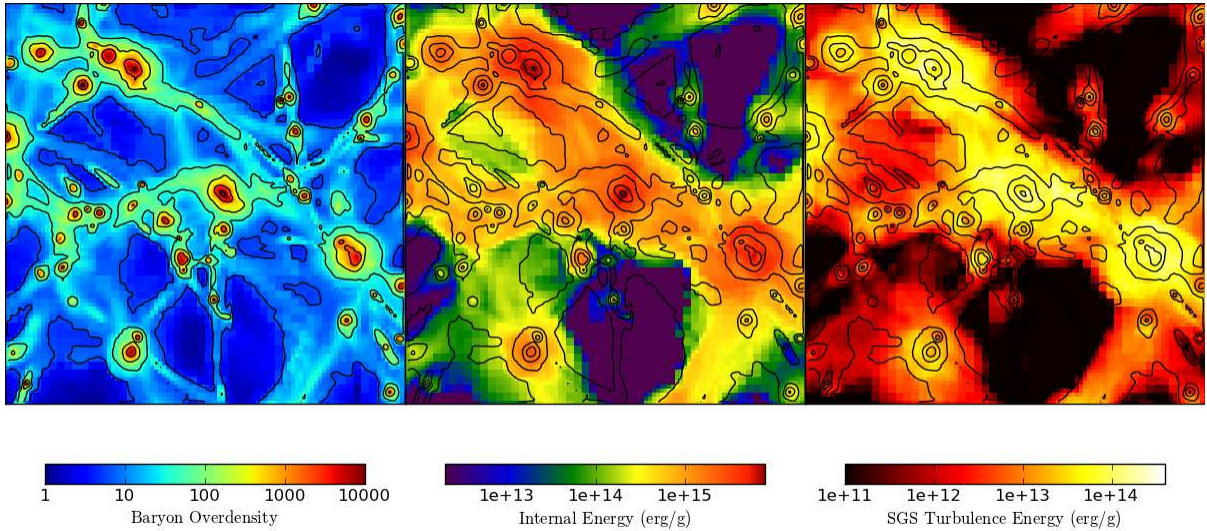}
\caption{The panels show projections of a cube with the side of $20\ \rmn{Mpc}\ h^{-1}$, extracted from the computational domain of the {\sc fearless} run, at $z = 0$. Baryon overdensity is shown at the left-hand side, internal energy in the central panel, and the SGS turbulent energy at the right-hand side, with density contours overlayed, and the corresponding colour bar under the panels.} 
\label{panels}
\end{figure*}

A first overview of the simulation data at $z = 0$ (Fig.~\ref{panels}) already shows a match of the locations where both the internal and the SGS turbulent energies are large. This indication is consistent with the close link occurring between gravitational collapse, virialisation and injection of turbulent energy during cosmological structure formation. In this Section, we use the SGS turbulent energy (measuring turbulent velocity fluctuation at the numerical cutoff scale) in relation to the internal energy (measuring the temperature) of the gas as a diagnostic of the properties of turbulence. Although turbulence is produced on length scales larger than the grid cutoff scale, the local energy injection is imprinted on the SGS turbulent energy because of the energy transport through the turbulent cascade. 

\begin{figure}
  \resizebox{\hsize}{!}{\includegraphics{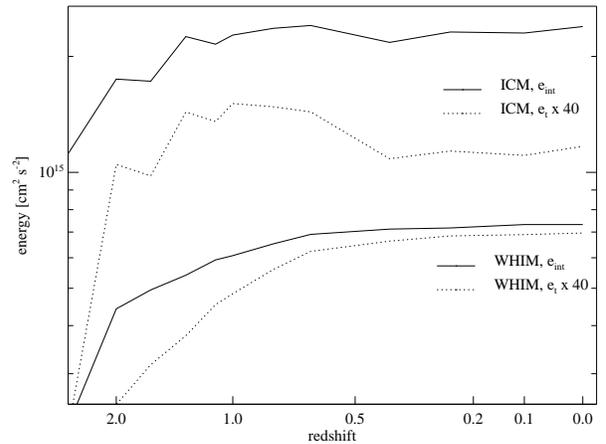}}
  \caption{Time evolution of the mass-weighted averages of specific internal ($e_{\rmn{int}}$, solid lines) and SGS turbulent ($e_{\rmn{t}}$, dotted lines) energies, for the two baryon phases under investigation. The two lines in the upper part of the plot refer to the ICM, and the other two to the WHIM. The lines are scaled according to the factors in the legends, in order to be accommodated in the same plot.} 
  \label{energy-z}
\end{figure}

In Fig.~\ref{energy-z} the temporal evolution of the mass-weighted average of the internal ($e_{\rmn{int}}$) and SGS turbulent ($e_{\rmn{t}}$) specific energies, both for the WHIM and ICM phase, is reported in more detail. From $e_{\rmn{t}}$ one can derive the average SGS turbulent velocities for the WHIM and ICM: they are $59$ and $76\ \mathrm{km\ s^{-1}}$, respectively, at $z = 0$. The corresponding average turbulent Mach numbers are 0.18 and 0.14 . 

The key feature to be noticed in Fig.~\ref{energy-z} is the different trend for the evolution of $e_{\rmn{t}}$ with time: for the WHIM phase, $e_{\rmn{t}}$ increases steadily to $z = 0$, whilst in the ICM it reaches a peak between $z \sim 1.0$ and $0.65$, and then decreases. In the following we speculate that this different evolution is related to the mechanisms of turbulence generation in the two baryon phases.  More specifically, the time evolution of $e_{\rmn{t}}$ in ICM and WHIM can be interpreted as turbulence production by mergers in the ICM, and by shock interactions in the WHIM.

There is supporting evidence corroborating our hypothesis. The gas in the ICM belongs to collapsed structures, which experienced merger episodes during their evolution. These merger events are related with the stirring of the baryons and the subsequent injection of turbulence (e.g., \citealt{mis09,pim10}). It is therefore not surprising that the maximum of $e_{\rmn{t}}$ in the ICM is consistent with the formation time (defined by the major merger phase) in the evolution of clusters in the mass range $10^{13}\ M_{\sun} < M < 10^{14}\ M_{\sun}$, as inferred by analytical models of hierarchical clustering (\citealt{lc93}, \citealt{st04}, and Fig.~4 of \citealt{gms07}). At later times, the level of SGS turbulence decreases, although the local injection in minor mergers likely contributes to slow down its decay (cf.~\citealt{ssh06}). 

As for the WHIM, the injection of turbulent energy at curved shocks is obviously related to the features of the gas accretion on filaments and haloes i.e., more specifically, on the amount of kinetic energy processed by the external shocks. In fact, a qualitative similarity can be noticed between the evolution of $e_{\rmn{t}}$ and the flux of kinetic energy through external shocks (Fig.~10 of \citealt{mrk00}, or Fig.~2 of \citealt{soh08}\footnote{The kinetic energy flux in \citet{mrk00} is averaged over all Mach numbers, whereas in \citet{soh08} it is sorted in Mach number, with results grossly comparable with the former study.}). This hints towards a link between the SGS turbulence energy production in the WHIM and the gas accretion on shocks associated with growing structures. Recently, \citet{clf10} investigated thoroughly the injection of turbulence in cluster outskirts (at the accretion shocks) with analytical calculations, finding that the turbulent support increases from $z \sim 0.5$. In their analysis, this is due to the shock weakening, the decrease of accretion rates on clusters and the decrease of the gas infall speed at low redshift. The evolution of $e_{\mathrm{t}}$ for the WHIM at redshift around 0.5 in Fig.~\ref{energy-z} is in qualitative agreement with the model of \citet{clf10}. 

In order to verify the resolution insensitivity of our results,
we analysed the innermost part of the {\sc fearless}
cluster simulation discussed in \citet{mis09}. In that setup, a
computational box with a side of $128\ \rmn{Mpc}\ h^{-1}$ is simulated
with a root grid resolution of $128^3$ cells and $128^3$ N-body
particles, but in a small cube with a side of $32\ \rmn{Mpc}\ h^{-1}$,
an additional static grid is nested and seven AMR levels are allowed,
such that the local root grid resolution is equivalent to $256^3$ for
both the mesh and the N-body particles, and the effective spatial
resolution is $7.8\ \rmn{kpc}\ h^{-1}$. This small volume is centred
on a growing cluster and therefore is not representative of a random
realisation of the cosmological initial conditions, but interestingly
in this region the time evolution of $e_{\rmn{t}}$ and $e_{\rmn{int}}$
is equal to that shown in Fig.~\ref{energy-z}. Our results therefore
look robust with respect to an increase of spatial and force
resolution.

\subsection{Compressive ratio}
\label{compr_ratio}

Our investigation, started from the analysis of the subgrid turbulent
energy in Fig.~\ref{energy-z}, is furthermore supported by the
structure of the velocity field at resolved scales. To this aim we
define the small-scale compressive ratio \citep{ko90,sfh09}: 
\begin{equation}
\label{compr-eq}
r_{\rmn{cs}} = \frac{\langle d^2 \rangle }{\langle d^2 \rangle  + \langle \omega^2 \rangle}
\end{equation}
where $\langle d^2 \rangle$ and $\langle \omega^2 \rangle$ are the averages of the squares of the divergence and the vorticity of the velocity field, respectively. This ratio quantifies the relative importance of compressional and solenoidal modes in a flow. As shown by \citet{sfh09}, the values of $r_{\rmn{cs}}$ tend to be higher for compressively-driven turbulence. 

\begin{figure}
  \resizebox{\hsize}{!}{\includegraphics{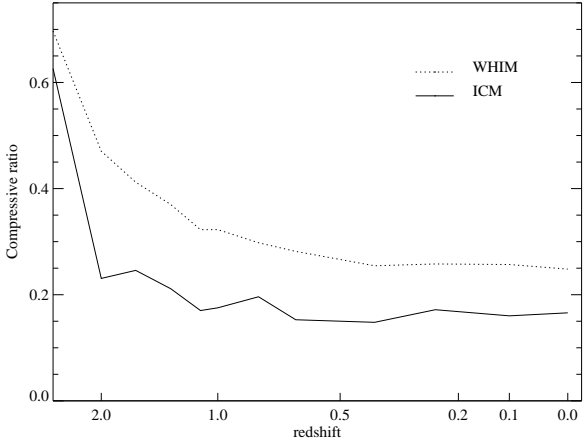}}
  \caption{Evolution of the mass-weighted compressive ratio $r_{\rmn{cs}}$ for the gas in the ICM (solid line) and WHIM (dotted line) temperature phase.} 
  \label{compr-ratio}
\end{figure}

\begin{figure}
  \resizebox{\hsize}{!}{\includegraphics{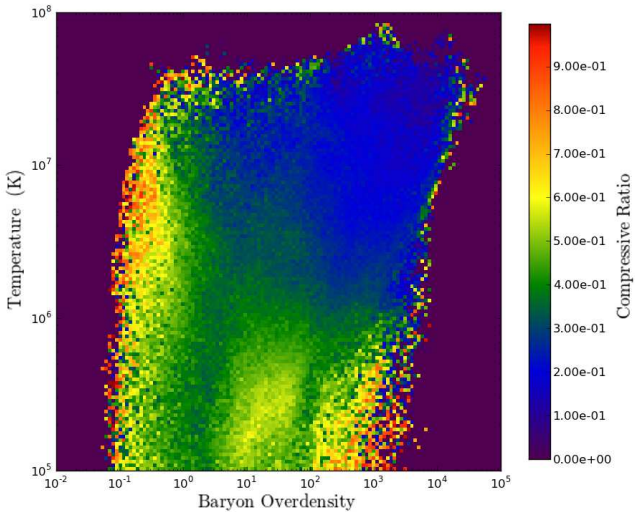}}
  \caption{Two-dimensional distribution function of the average compressive ratio $r_{\rmn{cs}}$ in the $T-\delta$ plane, at $z = 0$.} 
  \label{compr-ratio-t-rho}
\end{figure}

The evolution of $r_{\rmn{cs}}$ for the two baryon phases is reported in Fig.~\ref{compr-ratio}. As expected from the previous considerations, the gas in the WHIM phase has a larger value of the compressive ratio throughout the simulation, with respect to the ICM phase, indicating a higher contribution from compressional modes in filaments and cluster peripheries. A more detailed analysis on the $T-\delta$ plane at $z = 0$ (Fig.~\ref{compr-ratio-t-rho}) shows that the compressive ratio is low at high densities and temperatures (the ICM),
while it is significantly higher elsewhere. The compressive ratio is particularly high also at the extrema of the overdensity distribution, resulting either from strong rarefactions or compressions.

\begin{figure}
  \resizebox{\hsize}{!}{\includegraphics{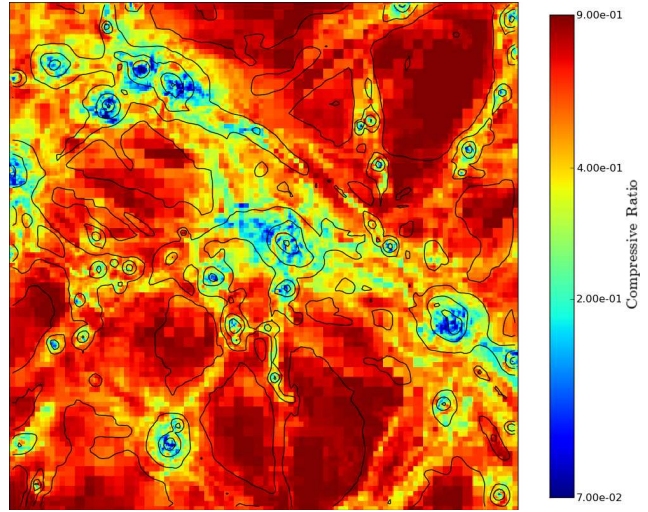}}
  \caption{Same projection volume as the panels in Fig.~\ref{panels}, but the compressive ratio $r_{\rmn{cs}}$ is shown. Baryon density contours are overlayed.} 
  \label{compr-projection}
\end{figure}

A similar conclusion can be drawn from the visual inspection of the projection in Fig.~\ref{compr-projection}: $r_{\rmn{cs}}$ is generally lower in clusters, except for localised regions (for example, in the cluster at the centre of the projected volume), likely to be associated with weak shocks in the ICM.

\subsection{Energy comparison with the adiabatic run}
\label{en-comp}

The role of the turbulence SGS model in our large-scale structure simulations and the consistency of the energy budget in this framework is further investigated with a comparison between the simulation using {\sc fearless} and an adiabatic reference run.
In these simulations the flow is mostly subsonic, and the energy content of the SGS turbulence is globally almost negligible \citep{mis09}. Quantitatively, it means that, for every baryon phase, the sum of kinetic and thermal energy should be approximately equal in the adiabatic and the {\sc fearless} run (in the latter case, the sum is extended to the SGS turbulent energy). 
In fact, good agreement (mostly within $2\%$) is found on the global gas properties of WHIM and ICM (mass fractions, energy content and their time evolution).

As a more sensitive diagnostic for the detailed energy budget in the ICM and WHIM, for the different phases we study the evolution of the total energies  
\begin{equation}
E_{\rmn{k}} = \sum_{i, phase} e_{\rmn{k},i} \rho_i V_i ,
\end{equation}
where $e_{\rmn{k},i}$ is the value of the specific energy $e_{\rmn{k}}$ in the cell $i$ (the index 'k' refers to $e_{\rmn{int}}$ or $e_{\rmn{t}}$; cf.~Fig.~\ref{energy-z}) and $\rho_i V_i$ is the baryon mass in the cell $i$. The sum is performed on the cells belonging to a same baryon phase, either WHIM or ICM.

\begin{figure}
  \resizebox{\hsize}{!}{\includegraphics{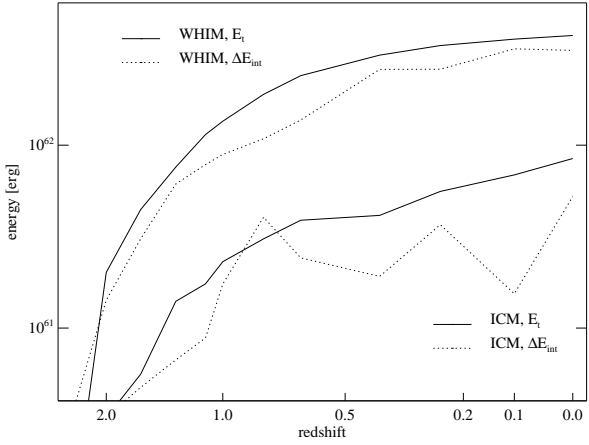}}
  \caption{Temporal evolution of the quantities $E_{\rmn{t}}$ (solid lines) and $\Delta E_{\rmn{int}}$ (dotted lines), defined in the text, for ICM and WHIM.} 
  \label{energy-x-omega}
\end{figure}

In Fig.~\ref{energy-x-omega} we compare $E_{\rmn{t}}$ with $\Delta E_{\rmn{int}}$, the difference of $E_{\rmn{int}}$ between the adiabatic and {\sc fearless} simulations. In both phases, $E_{\rmn{int}}$ (not in the plot) is 40 times or more larger than $E_{\rmn{t}}$. 

We observe that $E_{\rmn{t}}$ and $\Delta E_{\rmn{int}}$ are of the same order of magnitude during the simulation, both for the ICM and WHIM, indicating that the SGS turbulent energy acts as an energy buffer between the resolved and unresolved scales. In other words, the global decrease of $E_{\rmn{int}}$ in the {\sc fearless} run is partly balanced by $E_{\rmn{t}}$, so that the global energy budget is nearly unaffected. 
%On the other hand, especially for the ICM, $E_{\rmn{t}}$ is always larger than $\Delta E_{\rmn{int}}$, meaning that some fraction of $E_{\rmn{t}}$ was in the resolved kinetic energy of the adiabatic simulation. In both cases, the magnitude of $E_{\rmn{t}, i}$ is much smaller than $E_{\rmn{int}, i}$: on average, the SGS model provides a high-order correction to the energy budget of the simulation. As discussed in \citet{mis09}, local effects can nevertheless be important, and are more consistently taken into account in {\sc fearless}. 

Further physical interpretations of this energy budget are difficult, because of the high complexity of the flow in cosmological simulations; we refer the reader to \citet{mis09} to more tests in simplified setups. Another caveat, however, is that the turbulence SGS model is pushed to its limit in the WHIM because of the relatively high Mach numbers in the flow. For this reason, the result in the WHIM has to be confirmed with a SGS model that does not suffer from such constraints \citep[see][]{SchmFeder10}.

% On the contrary, in the WHIM phase, $E_{\rmn{t, WHIM}}$ is always larger than $\Delta E_{\rmn{int, WHIM}}$, meaning that the SGS turbulent energy mostly corresponds to a fraction of the the kinetic energy of the adiabatic simulation.
%The feature is reminiscent of the numerical tests performed in \citet[their Fig.~3]{mis09}, where it indicates a non-equilibrium state of the turbulent flow. If this interpretation is correct, turbulence in the ICM is in equilibrium between injection and dissipation, whereas this is not the case for the WHIM. This difference is again to be ascribed to the different stirring mechanisms in the two baryon phases.

\section{Thermal and turbulent pressure support}
\label{turb_press}

Using data from a cosmological hydrodynamic simulation, \citet{zff10} present an in-depth analysis of the vorticity and divergence fields in the intergalactic medium. The rationale behind their analysis is similar to ours, except that they infer turbulence properties from the derivative of the
resolved velocity field. Moreover, they consider dynamical equations for the modulus of the vorticity and the divergence. Of particular importance is the rate of change of the divergence, which is generalised to a co-moving coordinate system:
\begin{equation}
  \label{eq:zhudiv}
  \begin{split}
  \frac{\DD}{\DD t}d
  = &\frac{1}{a}\left[\frac{1}{2}\left(\omega^{2}-|S|^{2}\right) -
      \frac{1}{\rho}\nabla^{2}p + \frac{1}{\rho^{2}}\bmath{\nabla}\rho\cdot\bmath{\nabla} p \right] \\
    & - \frac{1}{a^2}\left[4\pi G (\rho + \rho_{\rm dm}) - \frac{3H^2}{2}\Omega_{\rm m}\right] - Hd,
  \end{split}
\end{equation}
where $a$ is the time-dependent cosmological scale factor, $\DD/\DD t=\partial/\partial t + a^{-1}\bmath{v}\cdot\bmath{\nabla}$ is the material derivative in co-moving coordinates,
$G$ is the gravitational constant, $\rho_{\rm dm}$ is the local dark matter density, $\Omega_{\rm m}$ is the cosmological mean density parameter of baryonic and dark matter, respectively, and $H=\dot{a}/a$ is the Hubble parameter. This is the same as in \citet{zff10} (their equation 3), with only the gravity terms (in the second line of equation \ref{eq:zhudiv}) slightly rearranged, and noting that, for the components of the rate of strain tensor, $2 S_{ij}S_{ij} = |S|^2$.  

The advantage of the filtering approach outlined in Section \ref{sgs} is that we can easily include SGS terms, in particular, turbulent pressure
terms. The SGS model described in Section \ref{sgs} allows for a direct computation of the turbulent pressure that is associated with the grid
scale: $p_{\rm t} = 2/3\ \rho e_{\mathrm{t}}$. Since the divergence equation is derived from the momentum equation, in which the turbulent pressure is simply added to the thermal pressure, it follows that the filtered version of the divergence equation is readily obtained from equation (\ref{eq:zhudiv}) by substituting $p$ with $p + p_{\mathrm t}$ everywhere:
\begin{equation}
  \label{eq:cosmodiv}
  \begin{split}
  \frac{\DD}{\DD t}d
  = &\frac{1}{a}\left[\frac{1}{2}\left(\omega^{2}-|S|^{2}\right) -
      \frac{1}{\rho}\nabla^{2}(p + p_{\mathrm t}) + \frac{1}{\rho^{2}}\bmath{\nabla}\rho\cdot\bmath{\nabla} (p + p_{\mathrm t}) \right] \\
    & - \frac{1}{a^2}\left[4\pi G (\rho + \rho_{\rm dm}) - \frac{3H^2}{2}\Omega_{\rm m}\right] - Hd,
  \end{split}
\end{equation}
and by considering filtered quantities (we dropped the hats 
for brevity). The trace-free part of the turbulence stress tensor is neglected in the above equation. The expression on the right-hand specifies the net negative compression rate of a fluid parcel. The self-gravity term on the very right stems from the Poisson equation for the gravitational potential and tends to decrease the divergence. 

To understand the meaning of the various terms in equation (\ref{eq:cosmodiv}), we consider different
limiting cases:
\begin{enumerate}
\item Incompressible limit: The fluctuations of the density with respect to the mean density vanish and $d=0$. Thus,
\begin{equation}
  \label{eq:div_incompress}
  \frac{1}{2}\left(\omega^{2}-|S|^{2}\right) = \frac{1}{\rho}\nabla^{2}(p + p_{\rm t}).
\end{equation}
\item Infinite resolution ($l_{\Delta}\rightarrow 0$): In this case, $p_{\rm t}$ vanishes, and equation~(\ref{eq:zhudiv}) is obtained.
\item Global filtering ($l_{\Delta}\sim L$, where $L$ is the integral length scale for turbulence injection): The filter formalism \citep{Germano1992} encompasses the limit of a statistical theory.
If flow structure is smoothed over the largest scales, i.~e., the size of galaxy clusters, the velocity derivative becomes negligible,
and the effect of turbulence is entirely given by the turbulent pressure:
\begin{equation}
\begin{split}
  \label{eq:divglobal}
  -&\frac{1}{\rho}\nabla^{2}(p + p_{\rm t}) + \frac{1}{\rho^{2}}\bmath{\nabla}\rho\cdot\bmath{\nabla}(p + p_{\rm t})=\\
  &\frac{1}{a}\left[4\pi G (\rho + \rho_{\rm dm}) - \frac{3H^2}{2}\Omega_{\rm m}\right].
\end{split}
\end{equation}
\end{enumerate}

Neglecting the effects of pressure gradients that are unaligned with the density gradients and comparing the limiting cases (ii) and (iii), we see that the term $1/2\ \rho(\omega^2-|S|^{2})$ in a fully resolved simulation is equivalent to $-\nabla^{2}p_{\rm t}$ if the flow is filtered on the largest scale of the system. In a large eddy simulation, we have an intermediate case, where part of the effect of turbulence is captured by the vorticity and the rate of strain of the resolved flow,
while the turbulent pressure at the grid scale accounts for numerically unresolved turbulence. 
If $\omega>|S|$, numerically resolved turbulence counteracts the gravitational contraction of the gas. 
The turbulent pressure of unresolved velocity fluctuations counteracts self-gravity if $\nabla^{2}p_{\rm t}<0$, respectively. The relative contribution of $p_{\rm t}$ depends on the grid scale. 

\begin{figure*}
\centering
\includegraphics[width=0.99\linewidth,clip]{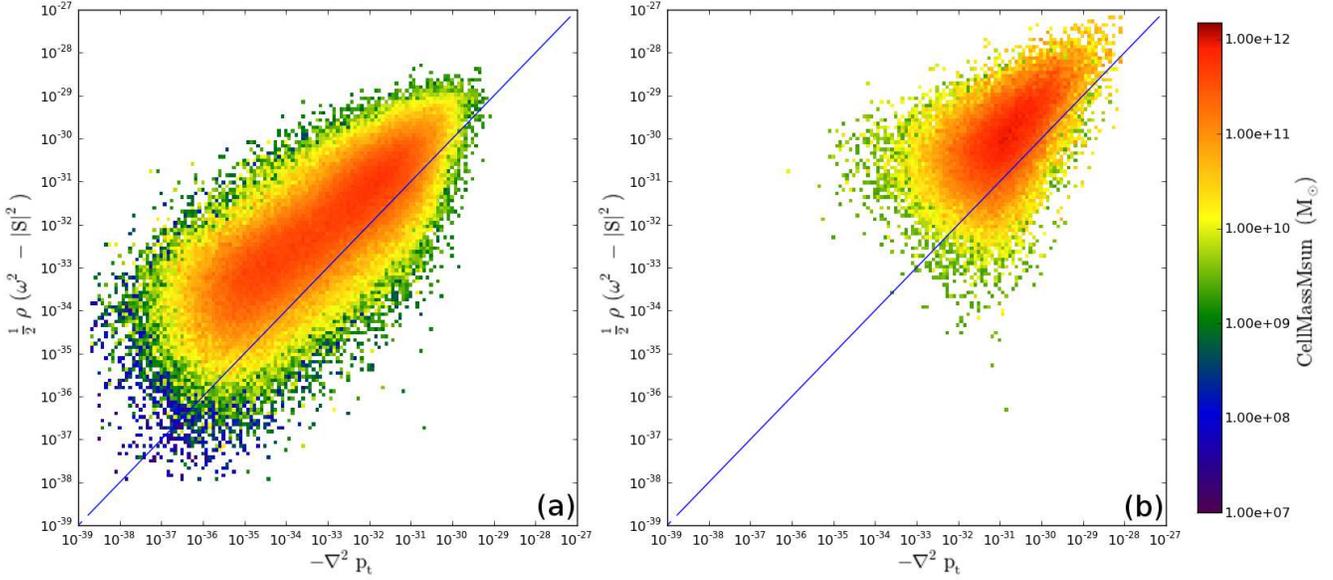}
\caption{Mass-weighted correlation diagrams of the Laplacians of the "resolved turbulent pressure", $1/2\ \rho(\omega^2-|S|^{2})$, and the turbulent pressure on subgrid scales, $-\nabla^{2}p_{\rm t}$, for the WHIM (a) and the ICM (b) at redshift $z=0$. In both panels, the diagonal line marks the location of the equality $1/2\ \rho(\omega^2-|S|^{2}) = -\nabla^{2}p_{\rm t}$.} 
\label{2dpdf-support}
\end{figure*}

It is important that, by its very definition, the turbulent pressure is a \emph{scale-dependent} quantity \citep{SchmFeder10}. \citet{zff10} investigate the scale-dependence of the turbulent pressure by integrating the spectrum of the kinetic energy density for all wave numbers greater than then a certain wave number (corresponding to a particular length scale). Since the resulting turbulent pressure spectrum is rather flat, no clear distinction is made between the integral turbulent pressure of the resolved flow and the turbulent pressure of velocity fluctuations below the grid scale. The advantage of our approach is that we can investigate both resolved turbulence and SGS turbulence effects. In Fig.~\ref{2dpdf-support}, we show the mass-weighted correlation diagrams of $1/2\ \rho(\omega^2-|S|^{2})$ and $-\nabla^{2}p_{\rm t}$. Both for the WHIM and the ICM, these quantities are roughly correlated. This is expected, because SGS turbulence is produced by the interactions with 
 turbulent velocity fluctuations on the smallest resolved length scales, which are probed by $\omega$ and $|S|$. However, the non-local nature of the SGS turbulence energy (see equation \ref{eq:etsum}), implies that there is no simple algebraic relationship between the resolved and unresolved turbulent pressures. This becomes manifest in the large scatter of the correlation diagrams. Consequently, the SGS model is essential for the computation of the support by the turbulent pressure, $-\nabla^{2}p_{\rm t}$.
Compared to a given value of the resolved turbulent pressure (corresponding to a horizontal cut through the two-dimensional distribution), $-\nabla^{2}p_{\rm t}$ is typically an order of magnitude smaller. Locally, however, the contribution from SGS turbulence can become comparable to or even exceed the resolved contribution, with the exception of extremely strong turbulence
 intensity in the upper right corner of the distribution ($1/2\ \rho(\omega^2-|S|^{2}) \ga 10^{-30}$ and $10^{-29}$, in arbitrary units, for panels (a) and (b) in Fig.~\ref{2dpdf-support}, respectively). 

\begin{figure*}
\centering
\includegraphics[width=0.99\linewidth,clip]{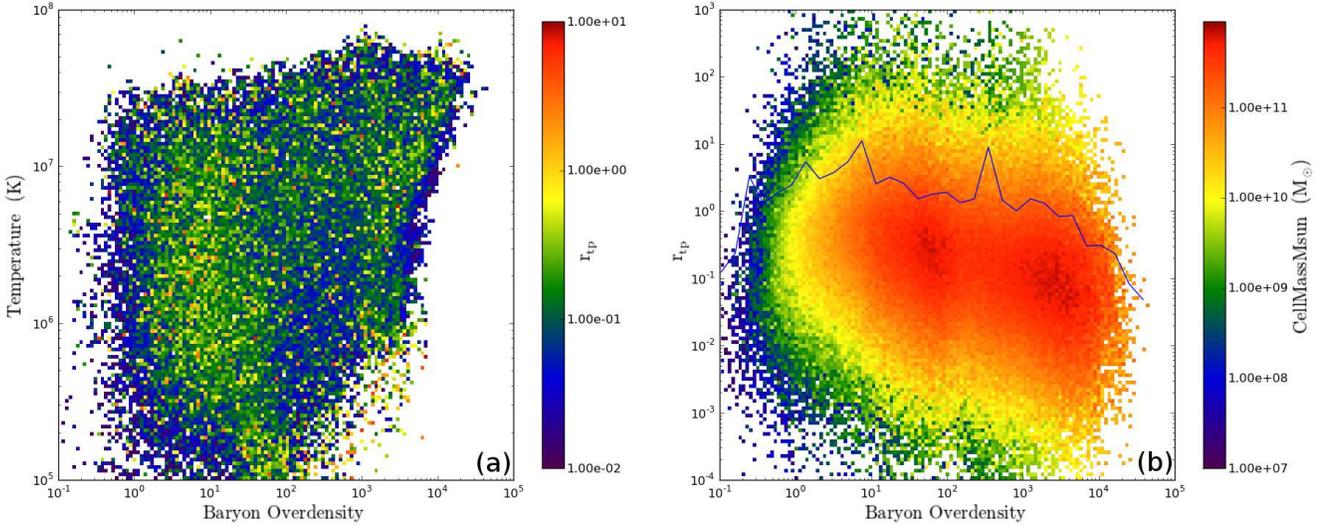}
\caption{Panel (a): two-dimensional distribution function of the average pressure ratio $r_{\mathrm{tp}}$, defined in equations (\ref{eq:ratio_turb_press}) and (\ref{constraints}), in the $T-\delta$ plane, , at $z=0$. Panel (b): two-dimensional mass distribution function of $r_{\mathrm{tp}}$ as a function of baryon overdensity. The mass-weighted average is overlayed.} 
\label{support-eq}
\end{figure*}

To analyse the effects of intense vorticity relative to the support of the gas due to its the thermal pressure, \citet{zff10} calculate the ratio of $1/2\ \rho(\omega^2-|S|^{2})$ to
$-\nabla^{2}p$.\footnote{However, $\nabla^{2}p$ appears  with a positive sign in figure 10 of \citet{zff10}. It is not clear whether this is an error in the labelling or an inconsistency of their calculation.} According to the above discussion, the full effect of turbulence is given by the sum of the vorticity term and the negative Laplacian of the turbulent pressure on the
grid scale. To estimate the importance of the turbulent relative to the thermal pressure support, we plot in the panels of Fig.~\ref{support-eq} the average in the $T-\delta$ plane and the two-dimensional mass distribution function of the ratio
\begin{equation}
	\label{eq:ratio_turb_press}
	r_{\rm tp}:=-\frac{\frac{1}{2}\rho(\omega^2-|S|^{2}) - \nabla^{2}p_{\rm t}}{\nabla^{2}p}\ ,
\end{equation}
subject to the constraints 
\begin{equation}
  \frac{1}{2}\rho(\omega^2-|S|^{2}) - \nabla^{2}p_{\rm t} > 0 \quad\mbox{and}\quad -\nabla^{2}p>0
\label{constraints}
\end{equation}
%with the baryon overdensity $\delta$. % = \rho / (\Omega_{\rmn b} \rho_{\rmn{cr}})$, where $\rho_{\rmn{cr}} = 3 H_0^2 / 8 \pi G$ is the critical density.

The distribution  in Fig.~\ref{support-eq}(a) suggests higher values of $r_{\rm tp}$ for overdensities $\delta \sim 10$. To be able to discern this trend more clearly, in Fig.~\ref{support-eq}(b) the mass distribution function of $r_{\rm tp}$ as a function of $\delta$ is shown.
A bimodal mass distribution becomes apparent, where the
low-density peak is a feature of the WHIM, and the
high-density peak is related to the ICM. 
%The overlap between the distributions possibly indicates that a
%simple temperature threshold is insufficient to separate the two phases. 
%Regardless, the high-density peak of the WHIM results in part
%from heated, shock-compressed gas, for which the temperature is close
%to the temperature threshold of $10^7\ {\rm K}$.  
Besides fluctuations, the average of $r_{\rm tp}$ decreases for increasing baryon overdensity. Clearly, the turbulent support (resolved plus unresolved) is important for the low-density gas in the WHIM ($1 \la \delta \la 100$). For this gas, the average $r_{\rm tp}$ is larger than for the WHIM at higher overdensity, and for the ICM.
This fact can be linked with the energy evolution seen in Fig.~\ref{energy-z}: the driving of large-scale turbulence in the ICM at $z=0$ has passed its maximum and is declining, while turbulence production in the WHIM is just saturating. Moreover, at low redshift $e_{\mathrm{t}}$ is a larger fraction of $e_{\mathrm{int}}$ in the WHIM, rather than in the ICM (according to $M_{\mathrm{t}}$ for the two phases, cf.~Section \ref{results}). In the ICM we observe a declining trend of $r_{\rm tp}$ towards large overdensities, meaning that the support for the densest gas is, on the average, mainly thermal.  

A more precise comparison to \citet[their figure 10]{zff10}
is not straightforward, because we use mass distribution functions in place of their scatter plot, and we include the contribution of SGS turbulence to the total pressure. Moreover, since the WHIM and the ICM are not clearly separated in their plot, the trend with the density is obscured. Anyway, the average values in Fig.~\ref{support-eq} are qualitatively similar to those of \citet{zff10}, making us confident of the consistency of these two analyses. 

\begin{table}
\caption{Mass (second column) and volume fractions (third column) of
  WHIM and ICM gas, selected according to the definitions (first
  column) introduced in Section \ref{turb_press}, at $z = 0$.}
\label{fractions}
\centering
\begin{tabular}{l|c|c}

\hline
 & mass fraction & volume fraction  \\
\hline
$\Omega_{\mathrm{WHIM}}$ & 0.315 & $1.83 \times 10^{-2}$ \\
$F_{\omega, \mathrm{WHIM}} $   & 0.287 & 0.221 \\
$F_{r_{\mathrm{tp}}, \mathrm{WHIM}}$   & $7.74 \times 10^{-2}$ & $5.95 \times 10^{-2}$\\
\hline
$\Omega_{\mathrm{ICM}}$  & $7.98 \times 10^{-2}$ & $3.53 \times 10^{-5}$ \\
$F_{\omega, \mathrm{ICM}} $    & 0.523 & 0.496\\
$F_{r_{\mathrm{tp}}, \mathrm{ICM}} $ & 0.128 & 0.117 \\
\hline
\end{tabular}
\end{table}

It is very instructive to consider the relations in equation (\ref{constraints}) separately. The first one defines the locations where the flow is turbulence-supported, regardless of the thermal pressure support on the divergence equation. We notice that this relation selects regions where the vorticity of the flow $\omega$ is large, thus recalling one of the outstanding features of turbulence.

In Table \ref{fractions} we report the mass fractions of the WHIM and ICM gas $\Omega_{\mathrm{WHIM}}$ and $\Omega_{\mathrm{ICM}}$, and the fractions of the computational volume occupied by this gas. Then, with $F_{\omega, \mathrm{WHIM}}$ we define the fraction of $\Omega_{\mathrm{WHIM}}$ where the first relation in equation (\ref{constraints}) is fulfilled, and the same for the ICM gas. Both for the WHIM and ICM, $F_\omega$ is a considerable fraction of the gas, in mass and volume. Moreover, one can see that the ratio between the volume and mass fraction is much smaller for $\Omega$ than for $F_\omega$, because the '$\omega$-selected' gas is much less clumped than the whole density-selected gas. This is in agreement with Fig.~\ref{support-eq}, where we observed that the turbulent support is smaller for the densest gas, in both phases. 

We also observe that  the volume fraction of $F_{\omega, \mathrm{ICM}}$ is rather similar to the volume filling factor for turbulence in the ICM estimated analytically by \citet{ssh06}, and computed in the simulations of \citet{in08}. The latter analysis was based on a simplified definition for characterising the turbulent flow based on its vorticity, therefore it is not surprising that it compares well with the present study based on the turbulent pressure support.

As interesting as it may be, the information brought by $F_\omega$ has to be complemented: given the tight link between turbulence injection and thermalisation, it is meaningful to consider the role of the turbulence only in combination with the thermal component of the energy budget. To this aim, in equation (\ref{eq:ratio_turb_press}) we defined the ratio $r_{\mathrm{tp}}$ between the dynamical and the thermal terms opposing gravitational collapse in the divergence equation (\ref{eq:cosmodiv}). Conservatively, we will assume that for $r_{\mathrm{tp}} > 0.1$ the role of the dynamical support in equation (\ref{eq:cosmodiv}) becomes non-negligible. According to this threshold, let $F_{r_{\mathrm{tp}}, \mathrm{WHIM}}$ be defined as the fraction of $\Omega_{\mathrm{WHIM}}$ where both relations of equation (\ref{constraints}) are fulfilled, and $r_{\mathrm{tp}} > 0.1$. An analogous quantity is defined for the ICM, and both are reported in Table \ref{fractions}, in mass and volume fractions at $z = 0$.

Although the mass and volume fractions expressed by $F_\omega$ are
relatively large, the turbulent support is non-negligible only for a
smaller subset of these cells. 
$F_{r_{\mathrm{tp}}}$ for the ICM is of the order of $10\%$ of the global mass and volume, and somewhat smaller for the WHIM. 
This result suggests that a significant non-thermal pressure
support, counteracting gravitational contraction, is 
a local behaviour of the cosmic flow, rather than a widespread feature.

\section{Conclusions}
\label{conclusions}

In this work we studied the evolution of the energy budget of the ICM
and WHIM in mesh-based hydrodynamical cosmological
simulations. Besides the internal and the resolved kinetic energy,
the {\sc fearless} 
method combining AMR and LES allows us to also follow the evolution of
the unresolved, SGS turbulent energy, defined according to the model
described in \citet{mis09}. 
Since an energy cascade sets in from the integral length scale down to
the (unresolved) Kolmogorov scale, the SGS turbulence contains
information on turbulent injection and evolution at larger, resolved
scales, with the advantage of being computed cell-wise and thus easily
accessible without further post-processing the resolved gas velocity
data.  

A first result of this work is the indication of a production
of turbulence with different properties in different baryon phases (Fig.~\ref{energy-z}). In the
ICM, the SGS turbulent energy peaks approximately at the expected
formation redshift of the haloes with mass $10^{13}$ -- $10^{14}\
M_{\sun}$, indicating a turbulence driving mechanism associated with
merger events. Indeed, the compressive ratio of the ICM baryon phase
is relatively low (Fig.~\ref{compr-ratio}), as expected in a flow
where the driving mechanism is dominated by shearing motions. 

In the WHIM phase, the SGS turbulent energy grows more smoothly
with time, hinting towards a different production mechanism. It is
straightforward to call for the role of shocks (in particular,
external shocks) in this process, because they enclose the WHIM gas in
filaments and outer cluster regions. Interestingly, the flux of
kinetic energy through the external shocks, as simulated by
\citet{mrk00} and \citet{soh08}, closely resembles the temporal
evolution of $e_{\rmn{t}}$. A similar trend has been also predicted
analytically in the cluster peripheries by \citet{clf10}. 
Further evidence is provided by the compressive ratio for the
WHIM, which is larger than the values found for the ICM and thus
indicates a flow dominated by a compressive driving mechanism.  The
energy content of SGS turbulence is larger for the ICM phase
(Fig.~\ref{energy-z}), but the relative importance with respect to the
internal energy is larger for the WHIM, as testified also from the
slightly larger turbulent Mach number (Section~\ref{results}).  

Some cautionary remarks are needed for a correct interpretation of the
results. First of all, this bimodality is not unambiguous. As known,
weak shocks are ubiquitous in the ICM
\citep{mrk00,rkh03,Pfrommer_MNRAS_2006,soh08,vbg09}, thus adding a
compressive contribution in that baryon phase, and similarly small
clumps move also along the filaments \citep{dmm06}, contributing to
solenoidal driving modes in the WHIM. Nevertheless, the mixing of
compressive and solenoidal modes is dominated by the former for the
WHIM and by the latter for the ICM. It would be interesting to explore
if and how further stirring mechanisms like AGN outflows 
change the simulated scenario. 

\citet{zff10} investigate the effects of the turbulent pressure on the
rate of change of the divergence. This quantity turned out to be
important for the clustering of the baryonic gas in the gravitational
wells of the dark matter. Following \citet{ias08} and \citet{sfh09},
we propose to utilise the negative growth rate of the divergence as
control variable for adaptive mesh refinement in cosmological
simulations. 

The study of the divergence equation (\ref{eq:cosmodiv}) has been
extended with the SGS turbulence modelling, finding that: 
\begin{itemize}
\item In general, the contribution of unresolved pressure to equation
  (\ref{eq:cosmodiv}) is not a dominant term, although locally it can
  become relevant (Fig.~\ref{2dpdf-support}); 
\item The turbulent support (defined by the first relation in equation
  \ref{constraints}) is largest for the WHIM gas in the overdensity
  range $1 \la \delta \la 100$, and tends to be small for the ICM
  gas ($\delta \ga 10^3$); 
\item A non-negligible mass fraction of the WHIM ($28.7\%$) and ICM
  ($52.3\%$) is characterised by a large vorticity, according to the
  criterion given by the first relation in equation
  (\ref{constraints}). However, in most of this gas the thermal
  support is not smaller than the dynamical one, so that the mass
  fractions of the gas where $r_{\mathrm{tp}} > 0.1$ are only $7.7\%$
  and $12.8\%$ of the WHIM and ICM at $z = 0$, respectively. 
\end{itemize}

For completeness, we 
note that at $z = 0$ in our simulation about $31\%$ of the gas is in
the WHIM phase, and $8\%$ in the ICM phase: most of the
turbulence-supported gas is thus in the WHIM phase, although the ICM
gas is more tightly related to observables. The emerging picture is
that of turbulence as a low-redshift (mostly $z < 0.5$) feature in the
WHIM, while in the ICM the main stirring epoch is slightly
earlier. The total mass fraction of baryons with $T > 10^5\
\mathrm{K}$ where the first relation in equation (\ref{constraints})
holds is $13.2\%$, in good agreement with the previous investigation
of \citet{zff10}. 

The fact that $F_\omega > F_{r_{\mathrm{tp}}}$ means that the
turbulence-supported gas is a substantial fraction of the cosmic
baryons, but  
the flow producing significant non-thermal support fill only small fractions of
space.  
Although this result reflects a typical feature of turbulence and can
be understood by considering that
turbulence injection is a by-product of thermalisation during the
structure growth, this is apparently at odds with recent observational
and theoretical claims of non-thermal pressure support, especially in
the cluster outskirts (e.g.~\citealt{kou10,lkn09,clf10}). From the
presented results, it seems that turbulence has a dynamical role only
in a volume fraction of about $6\%$ of the gas with $T > 10^5\
\mathrm{K}$. One possibility is that the spatial resolution of our
simulation is not adequate to model the flow close to the accretion
shock of growing clusters, which plays an important role in injecting
turbulence in the cluster outskirts \citep{clf10}. However, we tested
the results in Table \ref{fractions} in the well-resolved, innermost
region of the computational box presented in \citet{mis09}, obtaining
basically the same results (see also Section \ref{results}).  

We observe that the diagnostics used here for the study of the
turbulent support (pressure Laplacians) are different from other
studies, based on ratios between turbulent and internal energies (or,
equivalently, pressure ratios). Only in \citet{lkn09}, as far as we
are aware, the analysis is complemented by the computation of pressure
gradients, the very quantities that appear in the cluster mass
estimate \citep{rem06}. The discrepancy of the definitions could
explain why turbulence 'support' (in the sense of the pressure ratio)
is found large and widespread in many galaxy cluster simulations,
although the dynamical role elucidated in the present work is much
less significant. One could guess that diagnostics like the energy or
pressure ratios track the turbulent gas in a way similar to
$F_\omega$, rather than $F_{r_{\mathrm{tp}}}$.  

This work should be considered an
exploratory study, to be complemented by future 
simulations, possibly
including a more sophisticated treatment of additional physics
(ionisation background, radiative cooling, galactic winds). 
 However, the physical processes discussed in this work are governed
 primarily by gravity, thus the role of additional physics is not
 expected to change the scenario drastically
 (cf.~\citealt{krc07}). There is also room for improvements in the SGS
 modelling of turbulence, focusing in particular on
 compression-dominated and inhomogeneous turbulence
 \citep{schmidt09}. As mentioned above, a useful complement to this
 study will be a similar analysis of turbulent pressure and turbulence
 support in simulations of single clusters. This is left for future
 work.  

The study of turbulence in many branches of astrophysics has been
making progress in recent years, both from the theoretical and the
observational viewpoint. The more one goes into the details of this
field, the clearer it becomes that a simplistic way of approaching
turbulence (like simply assuming the classical reference
results by \citealt{k41} for energy spectra) is wrong or incomplete in
most cases. Turbulence driving already emerged as a key issue
in the context of compressible turbulence in isothermal gas, relevant
for the problem of star formation \citep{frk09}.  
\citet{KritUst10}, on the other hand, argue that a universal scaling
law should be observed at length scales that are sufficiently small
compared to the forcing scale. This idea is supported by recent
findings of 
\citet{SchmFeder10}. In this work, a subgrid scale model for
supersonic turbulence indicates a scaling 
exponent for the unresolved turbulent pressure that is independent of
the forcing. Nevertheless, the global statistics of turbulent pressure
varies with the forcing. The physical conditions of the gas and the
turbulence in the cosmological large-scale structure are obviously
different: here we are dealing mostly with subsonic or transonic
flow. As this study shows, one needs the investigation of the role of
turbulence forcing and turbulent support in this regime, and their
applications to cosmological simulations, in order to better
understand the gas dynamics in the IGM.  

\section*{acknowledgements}
The numerical simulations were carried out on the SGI Altix 4700 {\it
 HLRB-II} of the Leibniz Computing Centre in Garching (Germany). The {\sc enzo} code is developed by the Laboratory for Computational Astrophysics at the University of California in San Diego. The data analysis was performed using the {\tt yt} toolkit \citep{tso10}. Thanks to Carlo Giocoli for useful discussions, and to the anonymous referee for constructive comments, which improved the manuscript. 

\bibliography{cluster-index}
\bibliographystyle{bibtex/mn-web}

\appendix

\section[]{Different definitions of the baryon phases}
\label{resolution}

\begin{figure}
  \resizebox{\hsize}{!}{\includegraphics{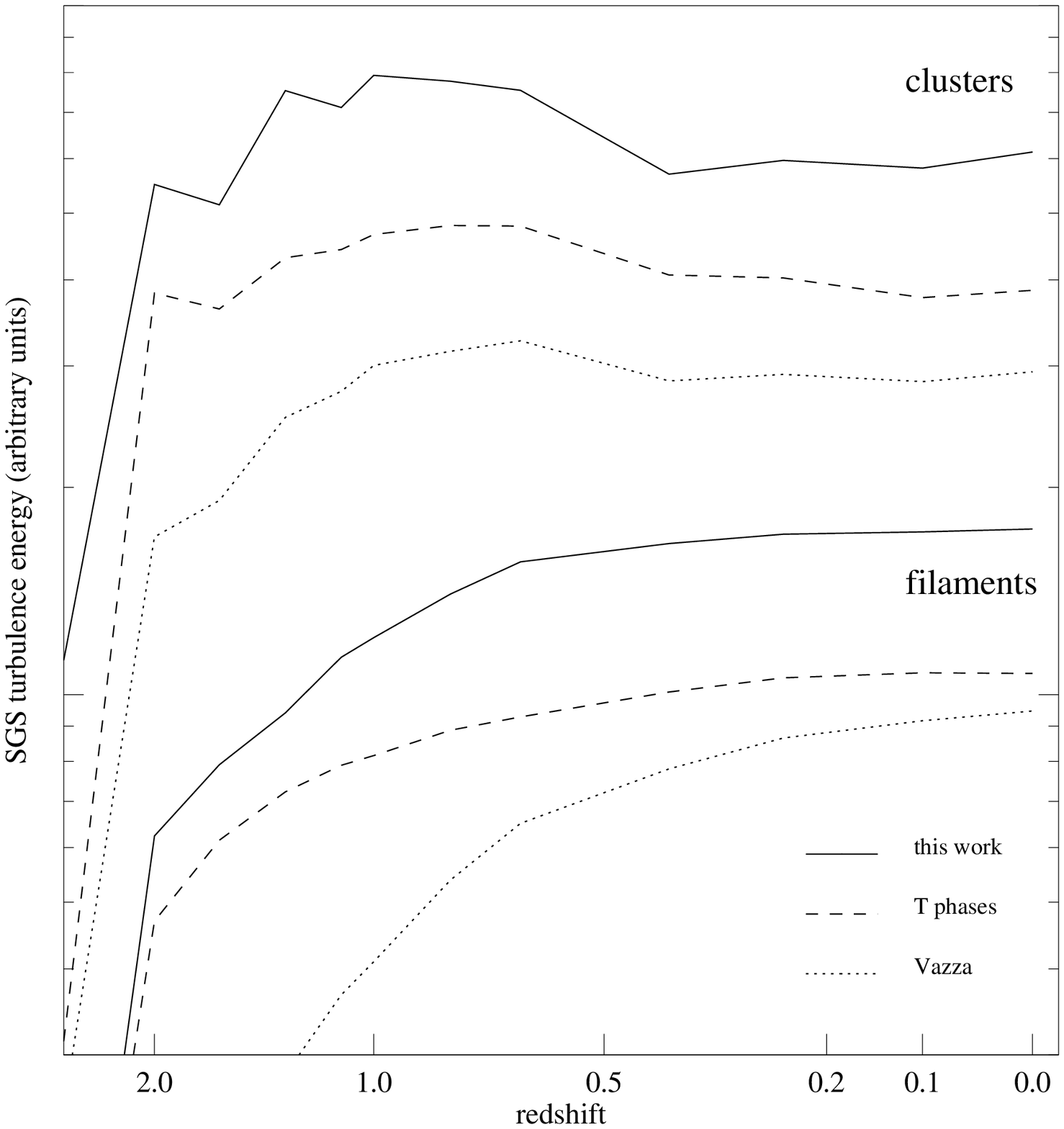}}
  \caption{Time evolution of the SGS turbulence energy for the phases labelled as `clusters' and `filaments' according to our definition based on baryon overdensity (solid lines) and on the criteria based on temperature (dashed lines) and on total overdensity \citep[][dotted lines]{vbg09}. The single lines are scaled by arbitrary factors, in order to ease the visualisation on the plot.} 
  \label{comparison-sgs}
\end{figure}

Throughout this work, the distinction between gas belonging to clusters or filaments has been based solely on baryon overdensity. We check here that the main results of this work do not depend on this definition by testing different characterisations of the baryons. In particular, we computed the evolution of the mass-weighted averages of the SGS turbulence energies of ``clusters'' and ``filaments'', according to criteria based on temperature \citep[e.g.,][]{co99} and on total overdensity \citep[e.g.,][]{vbg09}.

The temperature ranges defining the WHIM and ICM are $10^5\ \mathrm{K} < T < 10^7\ \mathrm{K}$ and $T > 10^7\ \mathrm{K}$, respectively.
In \citet{vbg09}, the baryon phases are characterised by the total overdensity $\delta_{\rmn t} = (\rho + \rho_{\rmn{dm}}) / \rho_{\rmn{cr}}$, where the suffix 'dm' indicates the dark matter density. 
In that work, filaments and sheets have overdensity $3 \leq \delta_{\rmn t} < 30$, and the clusters are defined by $\delta_{\rmn t} \geq 50$. 

 In Fig.~\ref{comparison-sgs} the evolution of $e_{\rmn t}$ for the two baryon phases and the different definitions is shown. The definitions are not completely equivalent and therefore the average values differ from each other, in some cases up to an order of magnitude, but in Fig.~\ref{comparison-sgs} they are scaled for sake of clarity. The comparison shows that the trends in the time evolution of $e_{\rmn t}$ presented and discussed in Section \ref{results} are apparent also with other definitions of the baryon phases, and are not caused by the particular choice of using baryon overdensity phases.

\bsp

\label{lastpage}

\end{document}